\documentclass[twocolumn]{aastex631}

\shorttitle{GRB\,221009A/SN\,2022xiw}
\shortauthors{Kong et al.}

\graphicspath{{./}{figures/}}

\begin{document}

\title{GRB\,221009A/SN\,2022xiw: A Supernova Obscured by a Gamma-Ray Burst Afterglow?}

\author{De-Feng Kong}
\affiliation{Guangxi Key Laboratory for Relativistic Astrophysics, School of Physical Science and Technology, Guangxi University, Nanning 530004, China; wangxg@gxu.edu.cn, lew@gxu.edu.cn}
\affiliation{GXU-NAOC Center for Astrophysics and Space Sciences, Nanning 530004, China}
\author{Xiang-Gao Wang}

\affiliation{Guangxi Key Laboratory for Relativistic Astrophysics, School of Physical Science and Technology, Guangxi University, Nanning 530004, China; wangxg@gxu.edu.cn, lew@gxu.edu.cn}
\affiliation{GXU-NAOC Center for Astrophysics and Space Sciences, Nanning 530004, China}
\author{WeiKang Zheng}
\affiliation{Department of Astronomy, University of California, Berkeley, CA 94720-3411, USA; weikang@berkeley.edu, afilippenko@berkeley.edu}
\author{Hou-Jun L\"{u}}
\affiliation{Guangxi Key Laboratory for Relativistic Astrophysics, School of Physical Science and Technology, Guangxi University, Nanning 530004, China; wangxg@gxu.edu.cn, lew@gxu.edu.cn}
\affiliation{GXU-NAOC Center for Astrophysics and Space Sciences, Nanning 530004, China}
\author{L. P. Xin}
\affiliation{CAS Key Laboratory of Space Astronomy and Technology, National Astronomical Observatories, Chinese Academy of Sciences, Beijing 100101, China.}
\author{Da-Bin Lin}
\affiliation{Guangxi Key Laboratory for Relativistic Astrophysics, School of Physical Science and Technology, Guangxi University, Nanning 530004, China; wangxg@gxu.edu.cn, lew@gxu.edu.cn}
\affiliation{GXU-NAOC Center for Astrophysics and Space Sciences, Nanning 530004, China}
\author{Jia-Xin Cao}
\affiliation{Guangxi Key Laboratory for Relativistic Astrophysics, School of Physical Science and Technology, Guangxi University, Nanning 530004, China; wangxg@gxu.edu.cn, lew@gxu.edu.cn}
\affiliation{GXU-NAOC Center for Astrophysics and Space Sciences, Nanning 530004, China}
\author{Ming-Xuan Lu}
\affiliation{Guangxi Key Laboratory for Relativistic Astrophysics, School of Physical Science and Technology, Guangxi University, Nanning 530004, China; wangxg@gxu.edu.cn, lew@gxu.edu.cn}
\affiliation{GXU-NAOC Center for Astrophysics and Space Sciences, Nanning 530004, China}

\author{B. Ren}
\affiliation{Guangxi Key Laboratory for Relativistic Astrophysics, School of Physical Science and Technology, Guangxi University, Nanning 530004, China; wangxg@gxu.edu.cn, lew@gxu.edu.cn}
\affiliation{GXU-NAOC Center for Astrophysics and Space Sciences, Nanning 530004, China}
\author{Edgar P. Vidal}
\affiliation{Department of Astronomy, University of California, Berkeley, CA 94720-3411, USA; weikang@berkeley.edu, afilippenko@berkeley.edu}
\author{ J. Y. Wei}
\affiliation{CAS Key Laboratory of Space Astronomy and Technology, National Astronomical Observatories, Chinese Academy of Sciences, Beijing 100101, China.}
\author{En-Wei Liang}
\affiliation{Guangxi Key Laboratory for Relativistic Astrophysics, School of Physical Science and Technology, Guangxi University, Nanning 530004, China; wangxg@gxu.edu.cn, lew@gxu.edu.cn}
\affiliation{GXU-NAOC Center for Astrophysics and Space Sciences, Nanning 530004, China}
\author{Alexei V. Filippenko}
\affiliation{Department of Astronomy, University of California, Berkeley, CA 94720-3411, USA; weikang@berkeley.edu, afilippenko@berkeley.edu}

\begin{abstract}

We present optical photometry for the afterglow of GRB\,221009A, in some respects the most extraordinary gamma-ray burst (GRB) ever observed.
Good quality in the $R$-band light curve is obtained, covering 0.32--19.57 days since the \textit{Fermi}-GBM trigger. We find that a weak bump emerges from the declining afterglow at $t \approx 11$ days; a supernova (SN) may be responsible. We use a smooth broken power-law and $^{56}\mathrm{Ni}$ model to fit the light curve. The best-fitting results reveal that the SN ejected a total mass of $M_\mathrm{ej} = 3.70\, M_\odot$, a $^{56}\mathrm{Ni}$ mass of $M_\mathrm{Ni} = 0.23 \, M_\odot$, and a kinetic energy of $E_\mathrm{SN,K} = 2.35 \times 10^{52}\,\mathrm{erg}$.
We also compare GRB\,221009A with other GRB-SN events based on a GRB-associated SN sample,
and find that only SN\,2003lw and SN\,2011kl can be obviously revealed in the afterglow of GRB\,221009A by setting these objects at its distance.
This suggests that a supernova (SN\,2022xiw) is possibly obscured by the brighter afterglow emission from GRB\,221009A.

\end{abstract}


\section{Introduction} \label{sec:intro}

The connection between long-duration gamma-ray bursts (LGRBs) and broad-line Type Ic supernovae (SNe~Ic-BL) has been established --- the ``GRB-SN connection'' \citep[e.g.,][]{2006ARA&A..44..507W,2012grb..book..169H,2017AdAst2017E...5C}.
Since the discovery of GRB\,980425/SN\,1998bw \citep{1998Natur.395..670G}, dozens of GRB-SNe have been observed, but relatively few were nearby (redshift $z<0.2$): GRB\,980425/SN\,1998bw \citep[$z=0.00867$;][]{1998Natur.395..670G}, GRB\,030329/SN\,2003dh \citep[$z=0.16867$;][]{2003ApJ...591L..17S}, GRB 031203/SN 2003lw \citep[$z=0.10536$;][]{2004ApJ...609L...5M}, GRB\,060218/SN\,2006aj \citep[$z=0.03342$;][]{2006Natur.442.1011P}, GRB\,100316D/SN\,2010bh \citep[$z=0.0592$;][]{2010arXiv1004.2262C}, GRB\,130702A/SN\,2013dx \citep[$z=0.145$;][]{2015A&A...577A.116D}, GRB\,161219B/SN\,2016jca \citep[$z=0.1475$;][]{2016GCN.20321....1T,2016GCN.20342....1D}, GRB\,171205A/SN\,2017iuk \citep[$z=0.0368$;][]{2019Natur.565..324I}, GRB\,180728A/SN\,2018fip \citep[$z=0.117$;][]{2018GCN.23142....1I}, and GRB\,190829A/SN\,2019oyw \citep[$z=0.0785$;][]{2019GCN.25565....1V}. 
However, some nearby LGRBs show no evidence of SN emission down to very deep limits: GRB\,060505  \citep[$z=0.089$;][]{2006Natur.444.1047F}, GRB\,060614 \citep[$z=0.1254$;][]{2006Natur.444.1047F,2006Natur.444.1050D}, and GRB\,111005A \citep[$z=0.01326$;][]{2018A&A...615A.136T,2018A&A...616A.169M}. Also, GRB\,211211A \citep[$z=0.0763$;][]{2022Natur.612..223R,2022Natur.612..228T} and the recent case of GRB\,230307A \citep[$z=0.065/3.87$;][which in terms of flux is only second to GRB\,221009A]{2024Natur.626..737L} have instead  been associated with potential kilonova emission. 

On 2022 October 9, at 13:16:59 (UTC dates are used herein), the \textit{Fermi} Gamma-Ray Burst Monitor \citep[GBM;][]{2009ApJ...702..791M} onboard the Fermi Gamma-ray Space Telescope (FGST) triggered an extraordinarily bright LGRB, GRB\,221009A \citep{2022GCN.32636....1V}. 
About 53~min later, at 14:10:17, the Burst Alert Telescope \citep[BAT;][] {2005SSRv..120..143B} onboard the \textit{Neil Gehrels Swift Observatory}
also triggered GRB\,221009A, but it was named Swift J1913.1+1946  \citep{2022GCN.32632....1D}. 
The \textit{Swift} X-Ray Telescope \citep[XRT;][]{2005SSRv..120..165B} began
observing the GRB at 14:13:09, 172\,s after the BAT trigger, and
the Ultraviolet and Optical Telescope \citep[UVOT;][]{2005SSRv..120...95R} began at 179\,s \citep{2023ApJ...946L..24W}.
GRB\,221009A is a nearby event, with a very low redshift $z=0.151$ \citep{2022GCN.32648....1D,2022GCN.32686....1C,2022GCN.32765....1I}. 
As GRB\,221009A optical afterglow observations continued, photometric evidence for an SN appeared \citep{2022GCN.32769....1B,2022GCN.32818....1B}.
A few days later,
\cite{2022GCN.32850....1M} announced the results of follow-up spectroscopy of the afterglow of GRB\,221009A, confirming the emerging contribution of SN\,2022xiw \citep{2022TNSCR3047....1P} as reported by \cite{2022GCN.32800....1D}. Moreover, \cite{2023ApJ...946L..22F} presented extensive optical photometry of the afterglow of GRB\,221009A, showing evidence for emission from an accompanying SN. \cite{2023ApJ...949L..39S} also presented their optical photometry and announced that they found an SN component.
However, \cite{2023ApJ...946L..25S} did not find SN signatures in their imaging and spectroscopy,
and \cite{2023ApJ...946L..28L} failed to see significant evidence for SN emission in their {\it James Webb Space Telescope (JWST)} and {\it Hubble Space Telescope (HST)} observations of the afterglow of GRB\,221009A. 
But late-time {\it JWST} observations possibly suggest some specific SN spectral features associated with GRB\,221009A. In particular, a close match with SNe Ic-BL suggests the presence of a typical GRB-SN in the spectrum \citep{2023GCN.33676....1B,2024NatAs.tmp...65B}.

Here we present our photometric follow-up observations of GRB\,221009A. We analyze the optical afterglow of GRB\,221009A by fitting the $R$-band light curve with a smooth broken power-law plus $^{56}\mathrm{Ni}$ model. In addition, we compare GRB\,221009A with other GRB-SN events based on a GRB-associated SN sample.
This paper is organized as follows. Section \ref{sec:Obervation and Data Reductions} presents our observations and data reduction, while Section \ref{sec:analysis and results} shows the analysis and results. Our conclusions and implications are presented in Section \ref{sec:Conclusion and discussions}.
Throughout, we adopt a concordance cosmology with parameters $H_0 = 69.3\, \mathrm{km\,s^{-1}\,Mpc^{-1}}$, $\Omega_\mathrm{M}=0.286$, $\Omega_\mathrm{\Lambda}=0.714$.

\section{Observations and Data Reduction} \label{sec:Obervation and Data Reductions}

The afterglow of GRB\,221009A was observed by many ground-based telescopes including our GWAC telescopes.
The GWAC system is an optical transient survey located at Xinglong Observatory, China; it includes two 60\,cm optical telescopes (GWAC-F60A/B), as one of the main ground-based facilities of the Space-based Multi-band Astronomical Variable Objects Monitor (SVOM)\footnote{SVOM is a China–France satellite mission dedicated to the detection and study of GRBs.} mission \citep{2016arXiv161006892W}.
GWAC-F60 began observing the afterglow of GRB\,221009A at 13:16:59, 22.04\,hr after the GBM trigger, with a set of $R$-band images.
In addition, we performed follow-up observations on October 14 and 16, but the optical afterglow was not detected in the stacked image. 
LCOGT \citep[Las Cumbres Observatory Global Telescope Network;][]{2013PASP..125.1031B} began observing GRB\,221009A about 7.73 hr after the GBM trigger; $R$-band images were obtained with the 1\,m Sinistro instrument at the Teide Observatory on Tenerife and the 1\,m Sinistro instrument at McDonald Observatory, Texas, USA.
$B$, $V$, $R$, and $I$ images of GRB\,221009A were also obtained with the 1\,m Nickel telescope at Lick Observatory \citep{2022GCN.32669....1V},
and additional $Clear$-band
images were obtained with the Lick 0.76\,m Katzman Automatic Imaging Telescope \citep[KAIT;][] {2001ASPC..246..121F}.

Point-spread-function (PSF) photometry was performed using DAOPHOT \citep{1987PASP...99..191S}
from the IDL Astronomy Users 
Library\footnote{\url{http://idlastro.gsfc.nasa.gov/}}.
Several nearby stars were chosen from the Pan-STARRS1\footnote{\url{http://archive.stsci.edu/panstarrs/search.php}} catalog for calibration;
their magnitudes were  transformed into the Landolt \citep{1992AJ....104..340L} magnitudes using the empirical prescription presented by Eq.~6 of \citet[][]{2012ApJ...750...99T}.

The photometry results are corrected for Galactic extinction with $E(B-V) = 1.36$\,mag \citep{2011ApJ...737..103S} for analysis. Owing to large uncertainties, we do not make corrections for the extinction in the GRB host galaxy. 
We report the original photometry from LCOGT, GWAC-F60, KAIT, and Nickel follow-up observations in Table \ref{tab:optical}. 

We collected additonal photometry data for our analysis from  \cite{2023ApJ...946L..24W}, \cite{2023ApJ...946L..25S}, \cite{2023ApJ...946L..23L}, \cite{2023ApJ...949L..39S} and Gamma-ray-burst Coordinates Network (GCN) Circulars \citep{2022GCN.32640....1B,2022GCN.32644....1H,2022GCN.32645....1B,2022GCN.32646....1D,2022GCN.32647....1X,2022GCN.32652....1B,2022GCN.32654....1D,2022GCN.32659....1P,2022GCN.32664....1R,2022GCN.32667....1C,2022GCN.32670....1K,2022GCN.32678....1G,2022GCN.32679....1R,2022GCN.32684....1B,2022GCN.32692....1W,2022GCN.32693....1S,2022GCN.32705....1B,2022GCN.32709....1V,2022GCN.32727....1M,2022GCN.32729....1Z,2022GCN.32730....1S,2022GCN.32739....1O,2022GCN.32743....1B,2022GCN.32750....1O,2022GCN.32752....1B,2022GCN.32753....1S,2022GCN.32755....1D,2022GCN.32758....1H,2022GCN.32759....1S,2022GCN.32765....1I,2022GCN.32769....1B,2022GCN.32795....1R,2022GCN.32803....1I,2022GCN.32804....1F,2022GCN.32809....1R,2022GCN.32811....1G,2022GCN.32818....1B,2022GCN.32852....1P,2022GCN.32860....1O,2022GCN.32934....1A}. XRT data were downloaded from the UK \emph{Swift} Science Data Center at the University of Leicester \citep{2009MNRAS.397.1177E} \footnote{\url{https://www.swift.ac.uk/xrt\_curves/01126853/}}. Figure \ref{fig:multi-band-PL} shows the multiband light curves of the afterglow in both optical and X-ray bands.

\startlongtable
\begin{deluxetable}{cccccc}
\centering
\tablecaption{Photometry of GRB\,221009A\tablenotemark{a} 
\label{tab:optical}}
\tablewidth{700pt}
\tabletypesize{\scriptsize}
\tablehead{
\colhead{$t_\mathrm{mid}$ (s)\tablenotemark{b}}&\colhead{$t_\mathrm{mid}$ (days)\tablenotemark{b}} & \colhead{Mag (Vega)\tablenotemark{c}} &\colhead{$1\sigma$}  & \colhead{Filter}  &  \colhead{Telescope} } 
\startdata
49088.07 	&	0.568 	&	17.04 	&	0.01 	&	$Clear$	&	KAIT	\\
49804.07 	&	0.576 	&	17.04 	&	0.01 	&	$Clear$	&	KAIT	\\
50516.09 	&	0.585 	&	17.09 	&	0.01 	&	$Clear$	&	KAIT	\\
54258.08 	&	0.628 	&	17.18 	&	0.01 	&	$Clear$	&	KAIT	\\
54970.10 	&	0.636 	&	17.28 	&	0.02 	&	$Clear$	&	KAIT	\\
55687.05 	&	0.645 	&	17.17 	&	0.02 	&	$Clear$	&	KAIT	\\
136147.05 	&	1.576 	&	18.36 	&	0.03 	&	$Clear$	&	KAIT	\\
136859.07 	&	1.584 	&	18.41 	&	0.02 	&	$Clear$	&	KAIT	\\
137576.07 	&	1.592 	&	18.42 	&	0.03 	&	$Clear$	&	KAIT	\\
140746.03 	&	1.629 	&	18.45 	&	0.04 	&	$Clear$	&	KAIT	\\
141390.06 	&	1.636 	&	18.53 	&	0.04 	&	$Clear$	&	KAIT	\\
142178.03 	&	1.646 	&	18.39 	&	0.04 	&	$Clear$	&	KAIT	\\
142986.04 	&	1.655 	&	18.25 	&	0.03 	&	$Clear$	&	KAIT	\\
143703.07 	&	1.663 	&	18.29 	&	0.05 	&	$Clear$	&	KAIT	\\
144419.07 	&	1.672 	&	18.19 	&	0.05 	&	$Clear$	&	KAIT	\\
229204.08 	&	2.653 	&	18.96 	&	0.05 	&	$Clear$	&	KAIT	\\
230275.09 	&	2.665 	&	19.15 	&	0.08 	&	$Clear$	&	KAIT	\\
314183.06 	&	3.636 	&	19.55 	&	0.05 	&	$Clear$	&	KAIT	\\
487790.04 	&	5.646 	&	20.02 	&	0.11 	&	$Clear$	&	KAIT	\\
574322.05 	&	6.647 	&	20.35 	&	0.14 	&	$Clear$	&	KAIT	\\
1005531.06 	&	11.638 	&	21.54 	&	0.51 	&	$Clear$	&	KAIT	\\
1349766.03 	&	15.622 	&	22.02 	&	0.86 	&	$Clear$	&	KAIT	\\
54648.09 	&	0.633 	&	19.13 	&	0.09 	&	$V$	&	Nickel	\\
55033.08 	&	0.637 	&	17.54 	&	0.02 	&	$R$	&	Nickel	\\
55588.03 	&	0.643 	&	15.99 	&	0.01 	&	$I$	&	Nickel	\\
59297.10 	&	0.686 	&	18.87 	&	0.06 	&	$V$	&	Nickel	\\
59631.03 	&	0.690 	&	17.62 	&	0.02 	&	$R$	&	Nickel	\\
59970.07 	&	0.694 	&	16.04 	&	0.01 	&	$I$	&	Nickel	\\
60305.04 	&	0.698 	&	20.61 	&	0.30 	&	$B$	&	Nickel	\\
60640.01 	&	0.702 	&	18.98 	&	0.07 	&	$V$	&	Nickel	\\
60974.04 	&	0.706 	&	17.64 	&	0.02 	&	$R$	&	Nickel	\\
61310.04 	&	0.710 	&	16.09 	&	0.01 	&	$I$	&	Nickel	\\
61645.02 	&	0.713 	&	20.05 	&	0.26 	&	$B$	&	Nickel	\\
61979.04 	&	0.717 	&	18.97 	&	0.08 	&	$V$	&	Nickel	\\
62314.10 	&	0.721 	&	17.64 	&	0.02 	&	$R$	&	Nickel	\\
62649.07 	&	0.725 	&	16.11 	&	0.01 	&	$I$	&	Nickel	\\
79414.90 	&	0.919 	&	18.15 	&	0.07 	&	$R$	&	GWAC-F60	\\
81203.40 	&	0.940 	&	18.47 	&	0.11 	&	$R$	&	GWAC-F60	\\
83394.80 	&	0.965 	&	18.11 	&	0.08 	&	$R$	&	GWAC-F60	\\
85059.10 	&	0.984 	&	18.22 	&	0.11 	&	$R$	&	GWAC-F60	\\
899095.30 	&	10.406 	&	18.07 	&	0.16 	&	$R$	&	GWAC-F60	\\
27964.40 	&	0.324 	&	16.52 	&	0.01 	&	$R$	&	LCOGT	\\
28290.60 	&	0.327 	&	16.55 	&	0.01 	&	$R$	&	LCOGT	\\
28617.50 	&	0.331 	&	16.56 	&	0.01 	&	$R$	&	LCOGT	\\
28945.80 	&	0.335 	&	16.58 	&	0.01 	&	$R$	&	LCOGT	\\
29272.80 	&	0.339 	&	16.57 	&	0.01 	&	$R$	&	LCOGT	\\
118637.60 	&	1.373 	&	18.59 	&	0.04 	&	$R$	&	LCOGT	\\
118964.30 	&	1.377 	&	18.54 	&	0.04 	&	$R$	&	LCOGT	\\
119293.20 	&	1.381 	&	18.56 	&	0.26 	&	$R$	&	LCOGT	\\
119620.10 	&	1.384 	&	18.56 	&	0.04 	&	$R$	&	LCOGT	\\
119947.60 	&	1.388 	&	18.58 	&	0.03 	&	$R$	&	LCOGT	\\
201327.50 	&	2.330 	&	19.52 	&	0.06 	&	$R$	&	LCOGT	\\
201654.90 	&	2.334 	&	19.59 	&	0.06 	&	$R$	&	LCOGT	\\
202309.00 	&	2.342 	&	19.61 	&	0.07 	&	$R$	&	LCOGT	\\
202636.30 	&	2.345 	&	19.47 	&	0.06 	&	$R$	&	LCOGT	\\
203159.70 	&	2.351 	&	19.53 	&	0.06 	&	$R$	&	LCOGT	\\
203487.30 	&	2.355 	&	19.46 	&	0.06 	&	$R$	&	LCOGT	\\
203814.90 	&	2.359 	&	19.55 	&	0.07 	&	$R$	&	LCOGT	\\
204142.10 	&	2.363 	&	19.41 	&	0.05 	&	$R$	&	LCOGT	\\
204469.90 	&	2.367 	&	19.36 	&	0.05 	&	$R$	&	LCOGT	\\
204912.50 	&	2.372 	&	19.60 	&	0.38 	&	$R$	&	LCOGT	\\
205566.30 	&	2.379 	&	19.49 	&	0.06 	&	$R$	&	LCOGT	\\
205893.20 	&	2.383 	&	19.49 	&	0.08 	&	$R$	&	LCOGT	\\
306667.70 	&	3.549 	&	20.00 	&	0.04 	&	$R$	&	LCOGT	\\
306994.10 	&	3.553 	&	20.15 	&	0.06 	&	$R$	&	LCOGT	\\
307323.60 	&	3.557 	&	20.24 	&	0.06 	&	$R$	&	LCOGT	\\
307651.10 	&	3.561 	&	20.20 	&	0.07 	&	$R$	&	LCOGT	\\
307978.00 	&	3.565 	&	19.98 	&	0.07 	&	$R$	&	LCOGT	\\
309028.90 	&	3.577 	&	20.06 	&	0.04 	&	$R$	&	LCOGT	\\
309355.70 	&	3.581 	&	20.06 	&	0.04 	&	$R$	&	LCOGT	\\
309682.80 	&	3.584 	&	20.04 	&	0.04 	&	$R$	&	LCOGT	\\
310010.00 	&	3.588 	&	20.03 	&	0.04 	&	$R$	&	LCOGT	\\
310336.70 	&	3.592 	&	20.09 	&	0.05 	&	$R$	&	LCOGT	\\
312204.39 	&	3.613 	&	20.10 	&	0.07 	&	$R$	&	LCOGT	\\
312530.89 	&	3.617 	&	20.13 	&	0.08 	&	$R$	&	LCOGT	\\
312857.29 	&	3.621 	&	20.08 	&	0.07 	&	$R$	&	LCOGT	\\
313184.64 	&	3.625 	&	20.18 	&	0.06 	&	$R$	&	LCOGT	\\
313512.20 	&	3.629 	&	20.16 	&	0.07 	&	$R$	&	LCOGT	\\
397818.80 	&	4.604 	&	20.41 	&	0.05 	&	$R$	&	LCOGT	\\
397859.82 	&	4.605 	&	20.42 	&	0.06 	&	$R$	&	LCOGT	\\
398146.40 	&	4.608 	&	20.45 	&	0.06 	&	$R$	&	LCOGT	\\
398186.69 	&	4.609 	&	20.51 	&	0.07 	&	$R$	&	LCOGT	\\
398474.00 	&	4.612 	&	20.64 	&	0.08 	&	$R$	&	LCOGT	\\
398514.09 	&	4.612 	&	20.49 	&	0.07 	&	$R$	&	LCOGT	\\
398803.10 	&	4.616 	&	20.62 	&	0.08 	&	$R$	&	LCOGT	\\
398841.59 	&	4.616 	&	20.44 	&	0.07 	&	$R$	&	LCOGT	\\
399129.60 	&	4.620 	&	20.62 	&	0.08 	&	$R$	&	LCOGT	\\
399170.09 	&	4.620 	&	20.44 	&	0.06 	&	$R$	&	LCOGT	\\
547867.00 	&	6.341 	&	20.95 	&	0.11 	&	$R$	&	LCOGT	\\
548194.30 	&	6.345 	&	20.84 	&	0.09 	&	$R$	&	LCOGT	\\
548522.80 	&	6.349 	&	20.87 	&	0.09 	&	$R$	&	LCOGT	\\
548849.70 	&	6.352 	&	20.87 	&	0.11 	&	$R$	&	LCOGT	\\
549178.30 	&	6.356 	&	21.07 	&	0.10 	&	$R$	&	LCOGT	\\
549614.70 	&	6.361 	&	20.79 	&	0.08 	&	$R$	&	LCOGT	\\
549942.50 	&	6.365 	&	20.84 	&	0.09 	&	$R$	&	LCOGT	\\
550269.00 	&	6.369 	&	20.96 	&	0.12 	&	$R$	&	LCOGT	\\
550596.20 	&	6.373 	&	20.92 	&	0.12 	&	$R$	&	LCOGT	\\
550922.90 	&	6.376 	&	20.86 	&	0.10 	&	$R$	&	LCOGT	\\
633664.60 	&	7.334 	&	21.12 	&	0.07 	&	$R$	&	LCOGT	\\
894065.30 	&	10.348 	&	21.74 	&	0.18 	&	$R$	&	LCOGT	\\
1236672.70 	&	14.313 	&	22.05 	&	0.16 	&	$R$	&	LCOGT	\\
1690581.70 	&	19.567 	&	22.94 	&	0.22 	&	$R$	&	LCOGT	\\
\enddata 
\tablenotetext{a}{Includes contributions from the GRB afterglow, host galaxy, and associated SN\,2022xiw}
\tablenotetext{b}{$t_\mathrm{mid}$ is the midpoint of each observation after the GBM trigger.}
\tablenotetext{c}{The data have not been corrected for extinction in the Milky Way Galaxy or the GRB host galaxy.}

\end{deluxetable}


\section{Analysis and Results} \label{sec:analysis and results}

\subsection{Modeling the Afterglow Light Curve}

In order to detect temporal features of the afterglow light curve, we fit it with a model having
two basic components: a smooth broken power-law (BPL) function and a $^{56}\mathrm{Ni}$ cascade decay model \citep[the $^{56}\mathrm{Ni}$ model; see detailed studies by, e.g.,][]{1979ApJ...230L..37A,1980ApJ...237..541A,1982ApJ...253..785A,1996snih.book.....A}. In addition, since the data are not corrected for host-galaxy emission, we use $\mathrm{F625W} = 24.88 \pm 0.08 \,\mathrm{mag}$ to correct for underlying host-galaxy light in our analysis, which approximately corresponds to the $R$ band and is measured by \cite{2023ApJ...946L..28L}.

The empirical BPL function
\citep[e.g.,][]{2007ApJ...670..565L,2012ApJ...758...27L,2015ApJS..219....9W} is given by
\begin{equation}
    F = F_0\left[\left(\frac{t}{t_\mathrm{b}}\right)^{\omega\alpha_1} + \left(\frac{t}{t_\mathrm{b}}\right)^{\omega\alpha_2}\right]^{-1/\omega},
\end{equation}
where $\alpha_1$ and $\alpha_2$ are the temporal slopes, $t_\mathrm{b}$ is the break time, and $\omega$ measures the sharpness of the break (in this paper, we fix $\omega = 3$).

For radioactivity from $^{56}\mathrm{Ni}$ and its daughter nucleus $^{56}\mathrm{Co}$, the total power can be written as \citep{2015ApJ...807..147W}
\begin{equation}
    P_\mathrm{Ni}(t) = \epsilon_\mathrm{Ni}M_\mathrm{Ni}e^{-t/t_\mathrm{Ni}} + \epsilon_\mathrm{Co}M_\mathrm{Ni}\frac{e^{-t/t_\mathrm{Co}}-e^{-t/t_\mathrm{Ni}}}{1-t_\mathrm{Ni}/t_\mathrm{Co}}\,\mathrm{erg\,s^{-1}},
\end{equation}
where $M_\mathrm{Ni}$ is the amount of $^{56}\mathrm{Ni}$ formed in the explosion, $\epsilon_\mathrm{Ni} = 3.9 \times 10^{10}\,\rm{erg\,g^{-1}\,s^{-1}}$, $t_\mathrm{Ni} = 8.8$\,days, $\epsilon_\mathrm{Co} = 6.8 \times 10^{9}\,\rm{erg\,g^{-1}\,s^{-1}}$, and $t_\mathrm{Co} = 111.3$\,days. 
The output luminosity can be written as \citep{1982ApJ...253..785A}
\begin{eqnarray}
    L_\mathrm{SN}(t) = e^{-(t/t_\mathrm{diff})^2}\int^t_0 2 P_\mathrm{Ni}(t^{'})\frac{t^{'}}{t_\mathrm{diff}}e^{(t^{'}/t_\mathrm{diff})^2}&
    \nonumber
    \\(1-e^{-At^{-2}})\frac{dt^{'}}{t_\mathrm{diff}},
\end{eqnarray}
where  
\begin{equation}
    t_\mathrm{diff} = \left(\frac{2\kappa M_\mathrm{ej}}{\beta cv_\mathrm{ej}}\right)^{1/2},
\end{equation}
is the diffusion time, and
\begin{equation}
    A = \frac{3\kappa_\gamma M_\mathrm{ej}}{4\pi v^2_\mathrm{ej}}
\end{equation}
is the leakage parameter \citep{2015ApJ...799..107W}.  
The parameter $\beta$ has a typical value of 13.8 \citep{1982ApJ...253..785A}; $M_\mathrm{ej}$, $v_\mathrm{ej}$, $\kappa$, $\kappa_\gamma$, and $c$ are the ejecta mass, the expansion velocity of the ejecta, the Thomson electron scattering opacity, the effective gamma-ray opacity, and the speed of light in a vacuum, respectively. 
We assume the velocity at the photosphere $v_\mathrm{phot} \approx v_\mathrm{ej}$. 
For a uniform density profile, the kinetic energy is given by $E_\mathrm{SN,K} = (3/10)\,M_\mathrm{ej} v^2_\mathrm{phot}$. 

We further assume that the spectral energy distribution (SED) in our SN model is a blackbody, which is a reasonable choice for SNe. The blackbody SED is calculated according to the Planck formula using the temperature and radius of the photosphere, implying the flux at frequency $\nu$ can be written as 
\begin{equation}
    f_\nu = \frac{2\pi h \nu^3}{c^2}\frac{1}{e^{h\nu/kT}-1}\,\rm{erg\,s^{-1}\,cm^{-2}\,Hz^{-1}},
\end{equation}
where $\nu$ is the frequency, $h$ is Planck’s constant, $k$ is Boltzmann’s constant, and $T$ is the temperature in degrees Kelvin. 

The temperature and radius are given by \citep{2017ApJ...850...55N}
\begin{equation}
    T_\mathrm{phot}(t) = \left \{ 
        \begin{array}{cc}
            \left( \frac{L_\mathrm{SN}(t)}{4 \pi \sigma v^{2}_\mathrm{phot} t^2} \right)^{\frac{1}{4}},  \quad  \left(\frac{L_\mathrm{SN}(t)}{4 \pi \sigma v^{2}_\mathrm{phot} t^2} \right)^{\frac{1}{4}}  > T_\mathrm{f} &  \\
            T_\mathrm{f},\qquad \qquad \left(\frac{L_\mathrm{SN}(t)}{4 \pi \sigma v^{2}_\mathrm{phot} t^2} \right)^{\frac{1}{4}} \leq  T_\mathrm{f} & 
        \end{array}
        \right.,
\end{equation}
\begin{equation}
    R_\mathrm{phot}(t) = \left\{ 
    \begin{array}{cc}
        v^{2}_\mathrm{phot}t , \qquad  \left(\frac{L_\mathrm{SN}(t)}{4 \pi \sigma v^{2}_\mathrm{phot} t^2} \right)^{\frac{1}{4}}  > T_\mathrm{f} &  \\
        \left(\frac{L_\mathrm{SN}(t)}{4 \pi \sigma T_\mathrm{f}^4} \right)^{\frac{1}{2}}, \quad \left(\frac{L_\mathrm{SN}(t)}{4 \pi \sigma v^{2}_\mathrm{phot} t^2} \right)^{\frac{1}{4}} \leq  T_\mathrm{f}& 
    \end{array}
    \right.,
\end{equation}
where $\sigma$ is the Stefan-Boltzmann constant and $T_\mathrm{f}$ is the final plateau temperature, an additional free parameter. This parameter simply allows us to extend our fits to later times, where other photospheric models based on determining the optical depth break down \citep{2013ApJ...770..128I}. 

\startlongtable
\begin{deluxetable}{llc}
\centering
\tablecaption{Model Parameters \label{tab:parameters}}
\tablewidth{700pt}
\tabletypesize{\scriptsize}
\tablehead{
\colhead{Parameter} & \colhead{Unit} &\colhead{Best fit} } 
\startdata
log\,$F_0$ &  $\mathrm{erg\,cm^{-2}\,s^{-1}}$          &   $-10.19$  \\
$\alpha_1$ &                                      &          0.75   \\
$\alpha_2$ &                                      &          1.56   \\
$t_\mathrm{b}$  & day                              &      0.44  \\
$\kappa$& $\rm{cm^2\,g^{-1}}$                  &       0.05          \\
log\,$\kappa_\gamma$ &$\rm{cm^2\,g^{-1}}$     &       $-0.65$      \\
$M_\mathrm{ej}$& $M_\odot$                             &       3.70       \\
$v_\mathrm{ej}$ & $10^9\mathrm{cm\,s^{-1}}$             &      3.26       \\
$M_\mathrm{Ni}$ & $M_\odot$                            &      0.40       \\
$T_\mathrm{f}$ & $10^3\mathrm{K}$                     &     3.50      \\
\enddata
\end{deluxetable}

\subsection{Light-Curve Fitting and Results}

\begin{figure}[ht!]
\centering
\includegraphics[angle=0,scale=0.17]{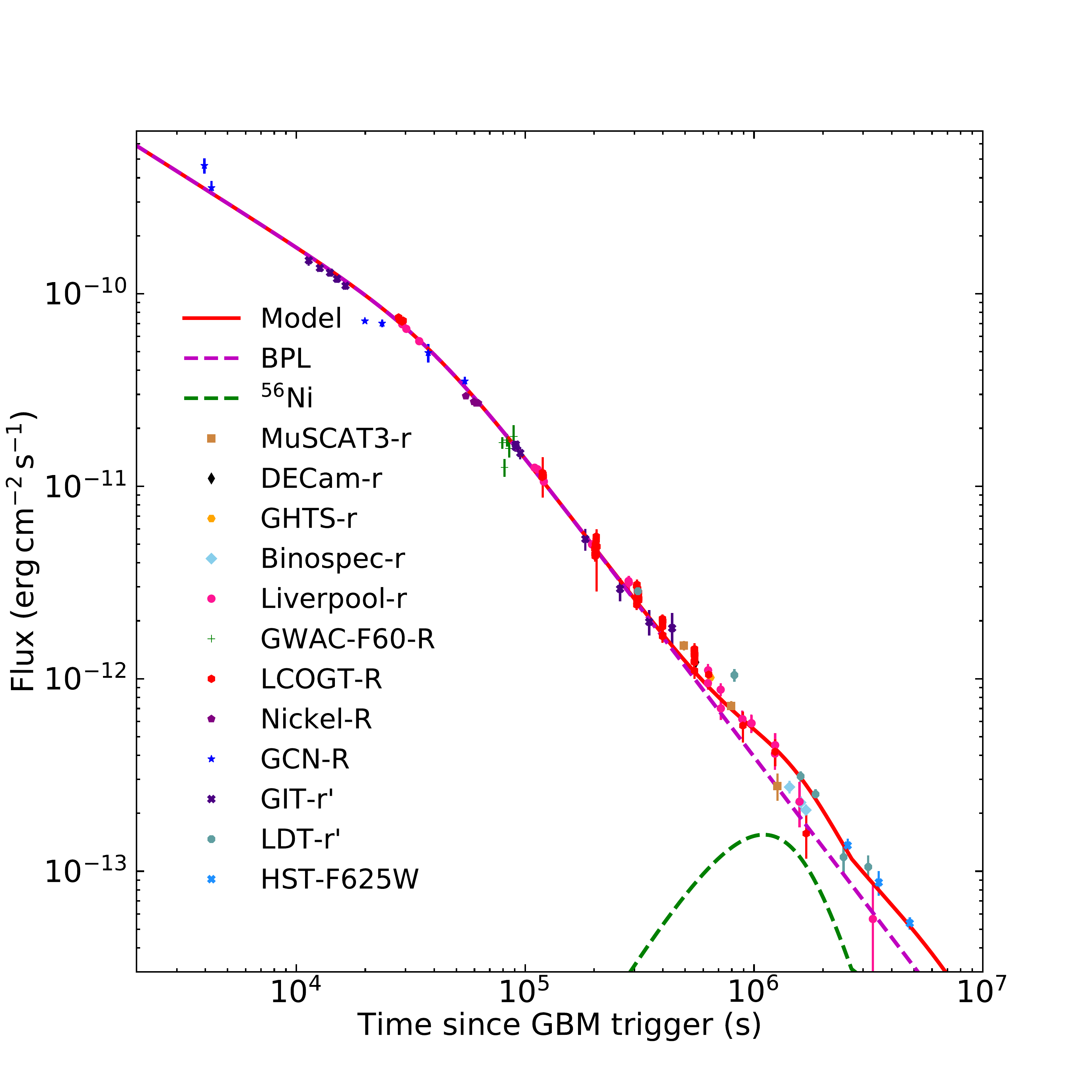}
\caption{$R$-band light curve of GRB\,221009A fitted with a BPL (purple dashed line) plus $^{56}\mathrm{Ni}$ model (green dashed line). The flux has been corrected for underlying host-galaxy light.
\label{fig:R-band-model}}
\end{figure}

For our light-curve analysis, we focus on the $R$ band because we have better data coverage in it compared with other bands. 
We used 10 parameters to fit the $R$ light curve; the best-fitting parameters are presented in Table \ref{tab:parameters} and the results are shown in Figure \ref{fig:R-band-model}. 

It should be noted that the model of\cite{1982ApJ...253..785A} that we adopt above might overestimate the $^{56}\mathrm{Ni}$ mass of SN\,2022xiw. Therefore, we calculate a more accurate value of the $^{56}\mathrm{Ni}$ mass using the equation derived by \cite{2019ApJ...878...56K},
\begin{eqnarray}
    M_\mathrm{Ni} &=& \frac{L_\mathrm{peak} \beta^{\prime2} t_\mathrm{peak}^2}{2 \epsilon t_\mathrm{Ni}^2}\Bigg( \left(1 - \frac{\epsilon_\mathrm{Co}}{\epsilon_\mathrm{Ni}}\right)  \nonumber \\ &\;&
      \times \left(1-\left(1+\beta^\prime t_\mathrm{peak}/t_\mathrm{Ni}\right)e^{-\beta^\prime t_\mathrm{peak}/t_\mathrm{Ni}}\right) \nonumber \\ &\;& 
    + \frac{\epsilon_\mathrm{Co}t_\mathrm{Co}^2}{\epsilon_\mathrm{Ni}t_\mathrm{Ni}^2}\left(1- \left(1 + \beta^\prime t_\mathrm{peak}/t_\mathrm{Co}\right) e^{-\beta^\prime t_\mathrm{peak}/t_\mathrm{Co}} \right)\Bigg)^{-1} \, , \label{newNi}
\end{eqnarray}
where we adopt a value of the mixing parameter $\beta^\prime = 0.56$, suitable for SNe~Ic-BL \citep{2021ApJ...918...89A}. Using the peak luminosity $L_\mathrm{peak}$ and the time of peak light $t_\mathrm{peak}$ of the bolometric light curve produced by the best-fitting parameters of the $R$-band light-curve fit, we find that the $^{56}\mathrm{Ni}$ mass of SN\,2022xiw is $0.23\,M_\odot$.

\begin{figure}[ht!]
\centering
\includegraphics[angle=0,scale=0.17]{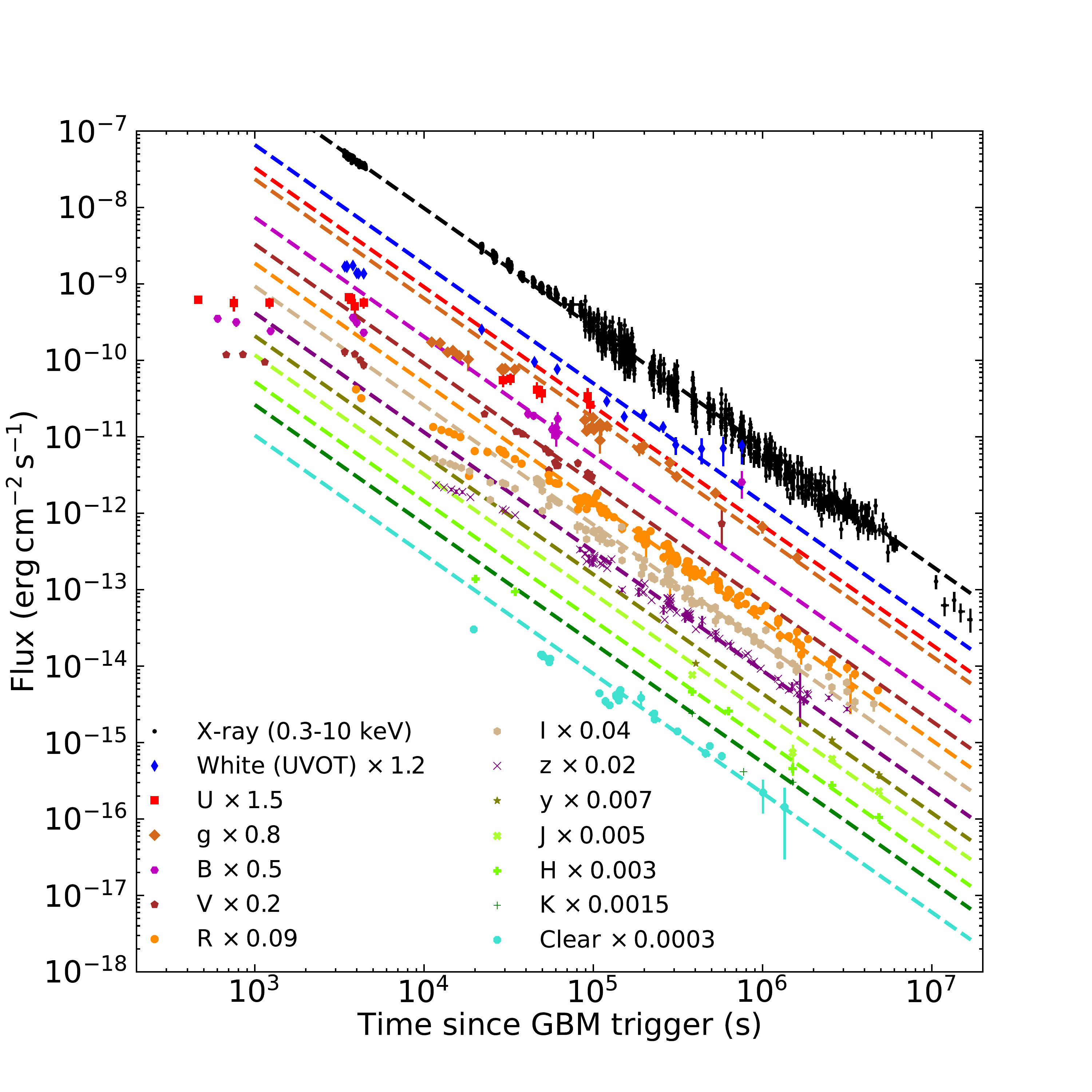}
\caption{Multiband light curves of GRB\,221009A described with a single power law, $f(t) \propto t^{-1.56}$, fit to all bands at late times.
\label{fig:multi-band-PL}}
\end{figure}

One can see from the Figure \ref{fig:R-band-model} that the bump is very weak, and the SN contribution is also small in the model. 
We note that the power-law slope of the later-time light curve is 1.56, and the X-ray power-law slope is also similar, $\sim 1.556$ \citep{2023ApJ...946L..22F}. We therefore tried to describe the multiband light curves with a single-decline-rate power law, $f(t) \propto t^{-1.56}$; the result is shown in Figure \ref{fig:multi-band-PL}, where
one can see that in all bands, there is no significant bump signature. 


\startlongtable
\begin{deluxetable*}{lllllll}
\centering
\tablecaption{Comparison of Parameters \label{tab:compare}}
\tablewidth{700pt}
\tabletypesize{\scriptsize}
\tablehead{
\colhead{Parameter} & \colhead{Unit} &\colhead{Fulton+ (2023)\tablenotemark{a}}  & \colhead{Srinivasaragava+ (2023)\tablenotemark{b}}& \colhead{Blanchard+ (2024)\tablenotemark{c}} & \colhead{Cano+ (2017)\tablenotemark{d}}  &  \colhead{Our result} } 
\startdata
$E_\mathrm{SN,K}$ & $10^{52}\,\mathrm{erg}$ &    2.6--9.0    &    1.6--5.2    &   --   & $2.52\pm1.79$    &   2.35   \\
$M_\mathrm{ej}$ & $M_\odot$               &   $7.1_{-1.7}^{2.4}$  &   3.5--11.1 &  --  & $5.90\pm3.80$    &  3.70  \\
$M_\mathrm{Ni}$ & $M_\odot$             &  $1.0_{-0.4}^{0.6}$   &     0.05--0.25  &  0.09   & $0.37\pm0.20$    & 0.23   \\
$E_{\mathrm{SN,K}}/M_\mathrm{ej}$ &$10^{52}\,\mathrm{erg}\,M_\odot^{-1}$ &  0.37--1.27 & 0.46--0.47 & -- & 0.43 &  0.64 \\
$M_{\mathrm{Ni}}/M_\mathrm{ej}$ &            &   0.141           &   0.014--0.023  &   --   &0.063   &   0.065    \\
$v_\mathrm{ej}$ &$10^9\,\mathrm{cm\,s^{-1}}$      &$3.39_{-0.57}^{0.59}$ &  2.8 & -- &    $2.02\pm0.85$ & 3.26  \\
\enddata
\tablenotetext{a}{The data from \cite{2023ApJ...946L..22F}.}
\tablenotetext{b}{The data from \cite{2023ApJ...949L..39S}.}
\tablenotetext{c}{The data from \cite{2024NatAs.tmp...65B}.}
\tablenotetext{b}{The data from Table 1 (GRB ALL) of \cite{2017AdAst2017E...5C}.}
\end{deluxetable*}


Table \ref{tab:compare} compares the parameters with the results reported by other works \citep{2023ApJ...946L..22F,2023ApJ...949L..39S,2024NatAs.tmp...65B} and the general value of GRB-SNe \citep{2017AdAst2017E...5C}. Our results are close to those of \cite{2023ApJ...949L..39S}, and consistent with the average value inferred by \cite{2017AdAst2017E...5C} for other GRB-SNe except for the expansion velocity of the ejecta,  $v_\mathrm{ej}$. \cite{2023ApJ...949L..39S} assumed SN\,2022xiw has a  photospheric velocity comparable to that of  SN\,1998bw, $v_\mathrm{ph} = 2.8\times10^9\,\mathrm{cm\,s^{-1}}$, and the value of our velocity is between that of \cite{2023ApJ...949L..39S} and \cite{2023ApJ...946L..22F}.
In addition, a spectrum taken of the optical afterglow $\sim 8$\,days after the burst reported the possible existence of broad features with velocities slightly larger than those of SN\,1998bw \citep{2022GCN.32800....1D}. Thus, we think our results are reasonable.

\begin{figure}[ht!]
\centering
\includegraphics[angle=0,scale=0.3]{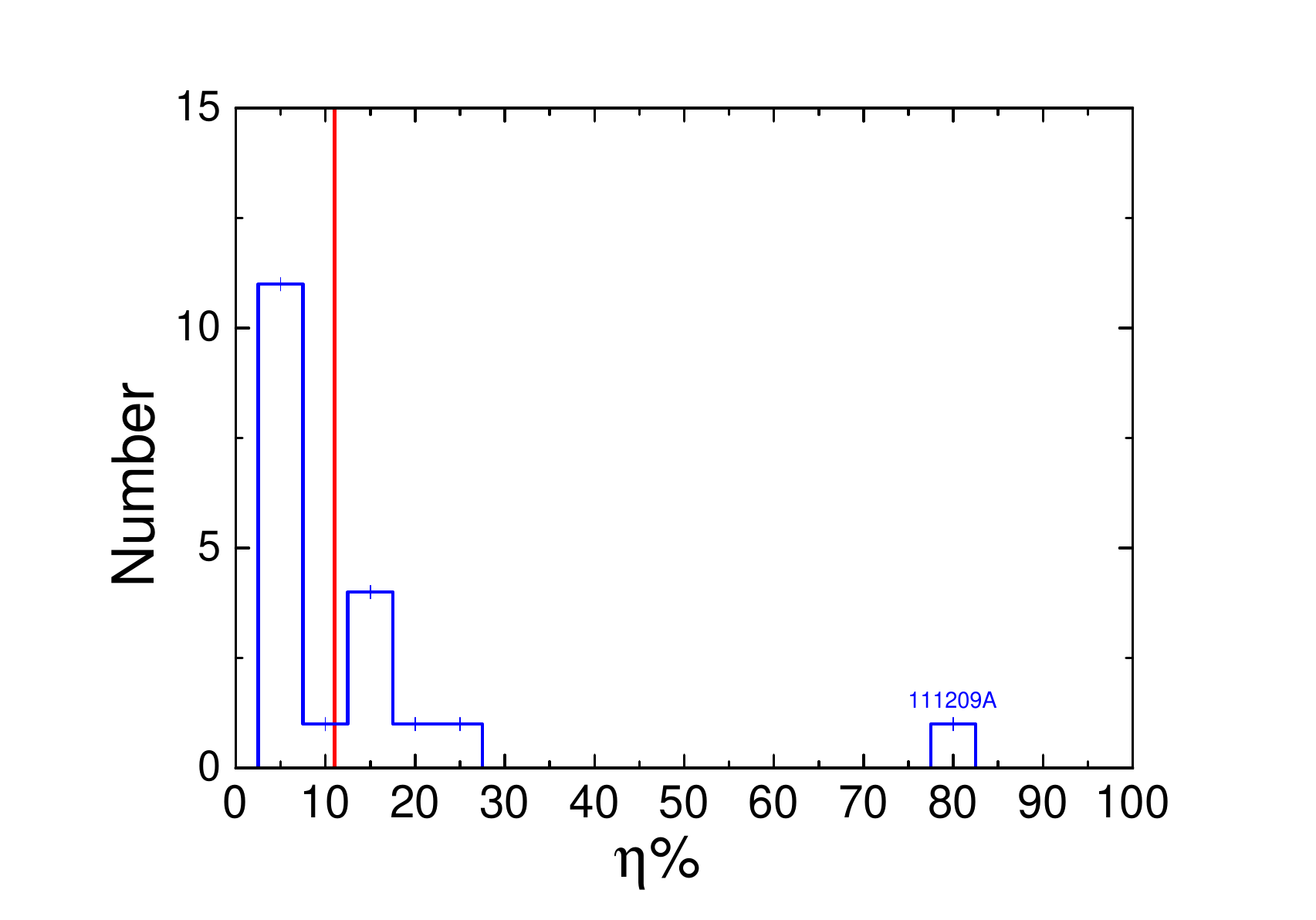}
\caption{Distribution of $\eta$ for GRB-SN events, reproduced from \cite{2018ApJ...862..130L}. The red solid line denotes the value for GRB\,221009A.
\label{fig:ratio}}
\end{figure}

One can also define the efficiency of GRB/SN events \citep{2018ApJ...862..130L}, 
\begin{equation}
    \eta = \frac{E_\mathrm{GRB}}{E_\mathrm{GRB} + E_\mathrm{SN,K}},
\end{equation}
to denote the energy partition. 
\cite{2022GCN.32762....1K} preliminarily estimated an isotropic energy release of log\,$E_\mathrm{\gamma,iso} = 54.77$ (0.1\,keV--100\,keV); \cite{2023ApJ...946L..23L} reported a beaming-corrected kinetic energy $E_\mathrm{\gamma,K} = 4\times10^{50}$\,erg and a small jet opening angle $\theta_{\rm jet}=1.64_{-0.20}^{+0.28}$. We therefore calculated the GRB energy $E_\mathrm{\gamma} = E_\mathrm{\gamma,iso}f_\mathrm{b} = 2.41\times 10^{51}\,\mathrm{erg}$ and $E_\mathrm{GRB}=E_\mathrm{\gamma}+E_\mathrm{\gamma,K}= 2.81\times 10^{51}\,\mathrm{erg}$, and get $\eta = 0.11$ for GRB\,221009A. 
Figure \ref{fig:ratio} shows the distribution of $\eta$ for the GRB-SN events; the $\eta$ value of GRB\,221009A is consistent with the majority of GRB-SNe \citep[the center value of $\eta \approx 0.1$;][]{2018ApJ...862..130L}.

\subsection{Comparison of GRB\,221009A and other GRB-SNe}

To better study the properties of GRB\,221009A, we compared GRB\,221009A with a sample of 14 other GRB-SN events ---
980425/1998bw, 030329/2003dh, 031203/2003lw, 060218/2006aj, 080109/2008d, 081007/2008hw, 091127/2009nz, 100316D/2010bh, 101219B/2010ma, 111209A/2011kl, 120422A/2012bz, 130427A/2013cq, 130702A/2013dx, 161219B/2016jca; their properties and parameters were obtained from \cite{2018ApJ...862..130L}. All GRB-SN events in our sample have strong evidence confirming an SN associated with a GRB \citep[see more details in the review by][and references therein]{2018ApJ...862..130L}.  
 
\begin{figure}[ht!]
\centering
\includegraphics[angle=0,scale=0.3]{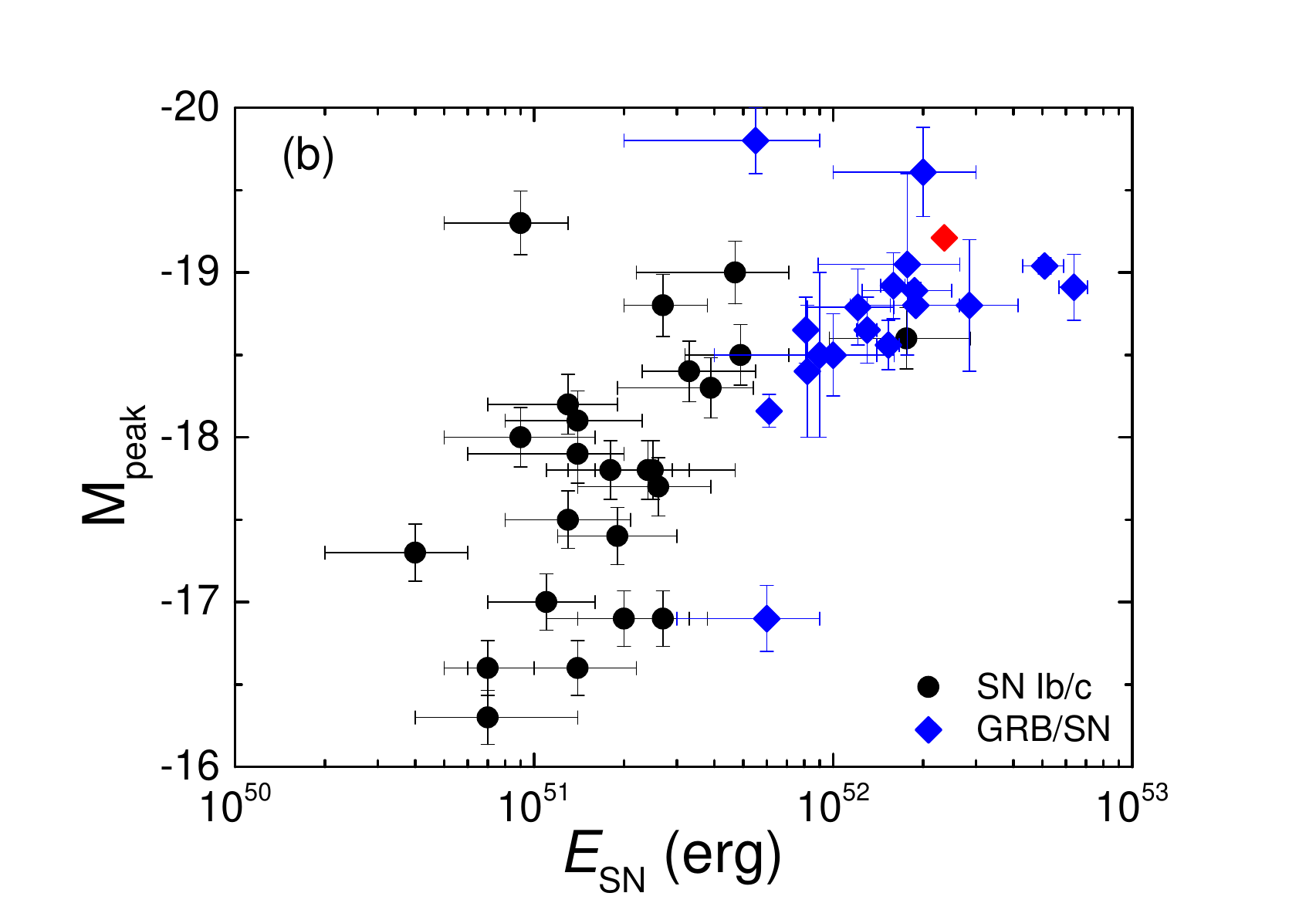}
\includegraphics[angle=0,scale=0.3]{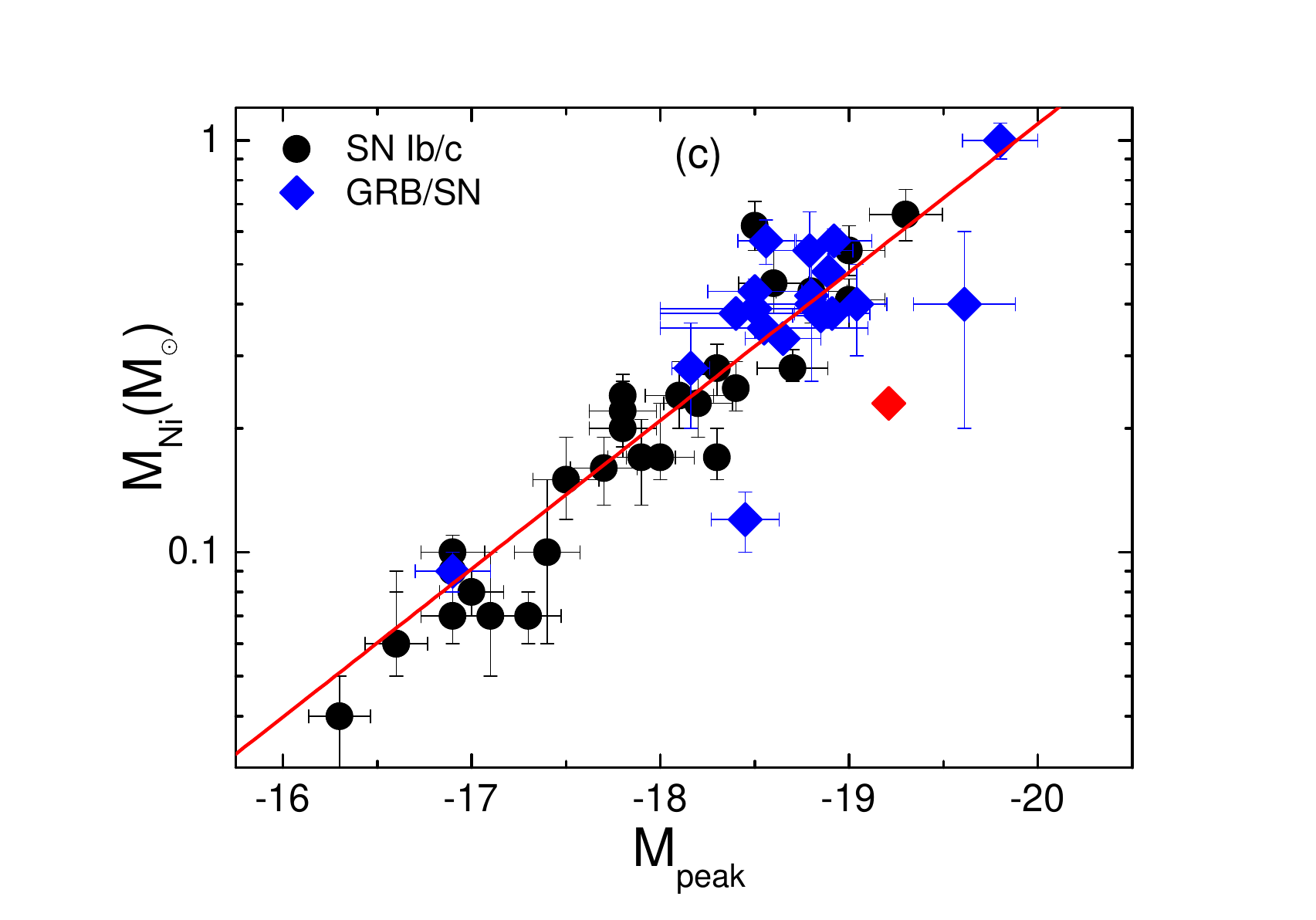}
\caption{Peak magnitude ($M_\mathrm{peak}$) of SNe as a function of $E_\mathrm{SN}$ {\it (top)} and $M_\mathrm{Ni}$ {\it (bottom)}. The red diamonds denote GRB\,221009A. Both panels are reproduced from \cite{2018ApJ...862..130L}.
\label{fig:E_sn-M_peak-M_Ni}}
\end{figure}

\begin{figure}[ht!]
\centering
\includegraphics[angle=0,scale=0.3]{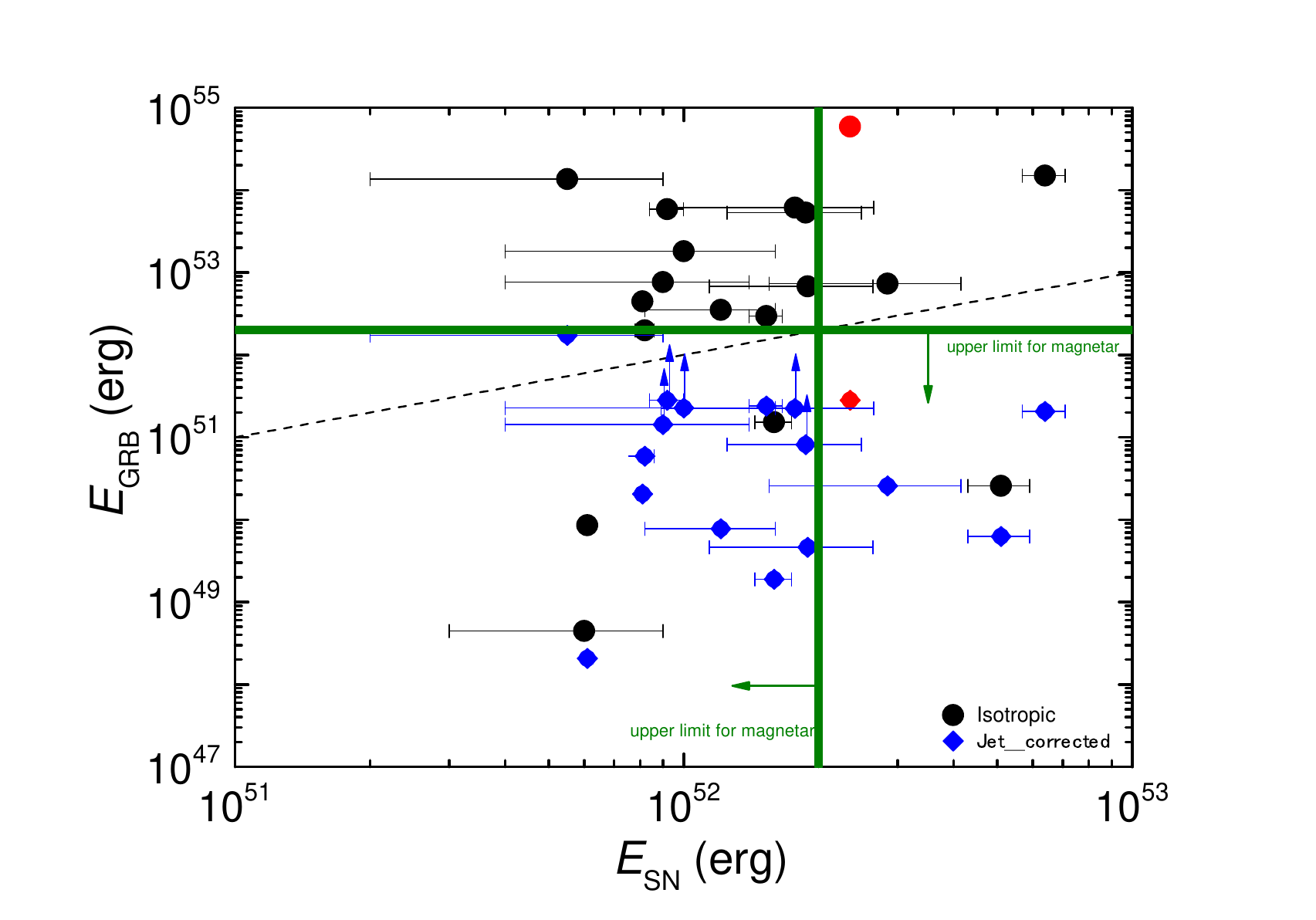}
\caption{$E_\mathrm{\gamma,iso}/E_\mathrm{GRB}$ vs. $E_\mathrm{SN}$. The vertical and horizontal lines are the upper limits of the magnetar energy budget. GRB\,221009A is represented with red points. The dashed line denotes the equality line. This figure is reproduced from \cite{2018ApJ...862..130L}.
\label{fig:E_grb-E_sn}}
\end{figure}

Figure \ref{fig:E_sn-M_peak-M_Ni} shows two correlations among SN parameters: peak magnitude ($M_\mathrm{peak}$) of the SN as a function of $E_\mathrm{SN}$ (top panel) and $M_\mathrm{Ni}$ (bottom panel). In both correlations, GRB\,221009A appears to be consistent with other GRB-SN events. 
Figure \ref{fig:E_grb-E_sn} shows $E_\mathrm{\gamma,iso}/E_\mathrm{GRB}$ as a function of $E_\mathrm{SN}$. One can see that both the $\gamma$-ray energy and the SN kinetic energy of GRB\,221009A are relatively large, which means that the SN (if present) is bright.

\begin{figure}[ht!]
\centering
\includegraphics[angle=0,scale=0.17]{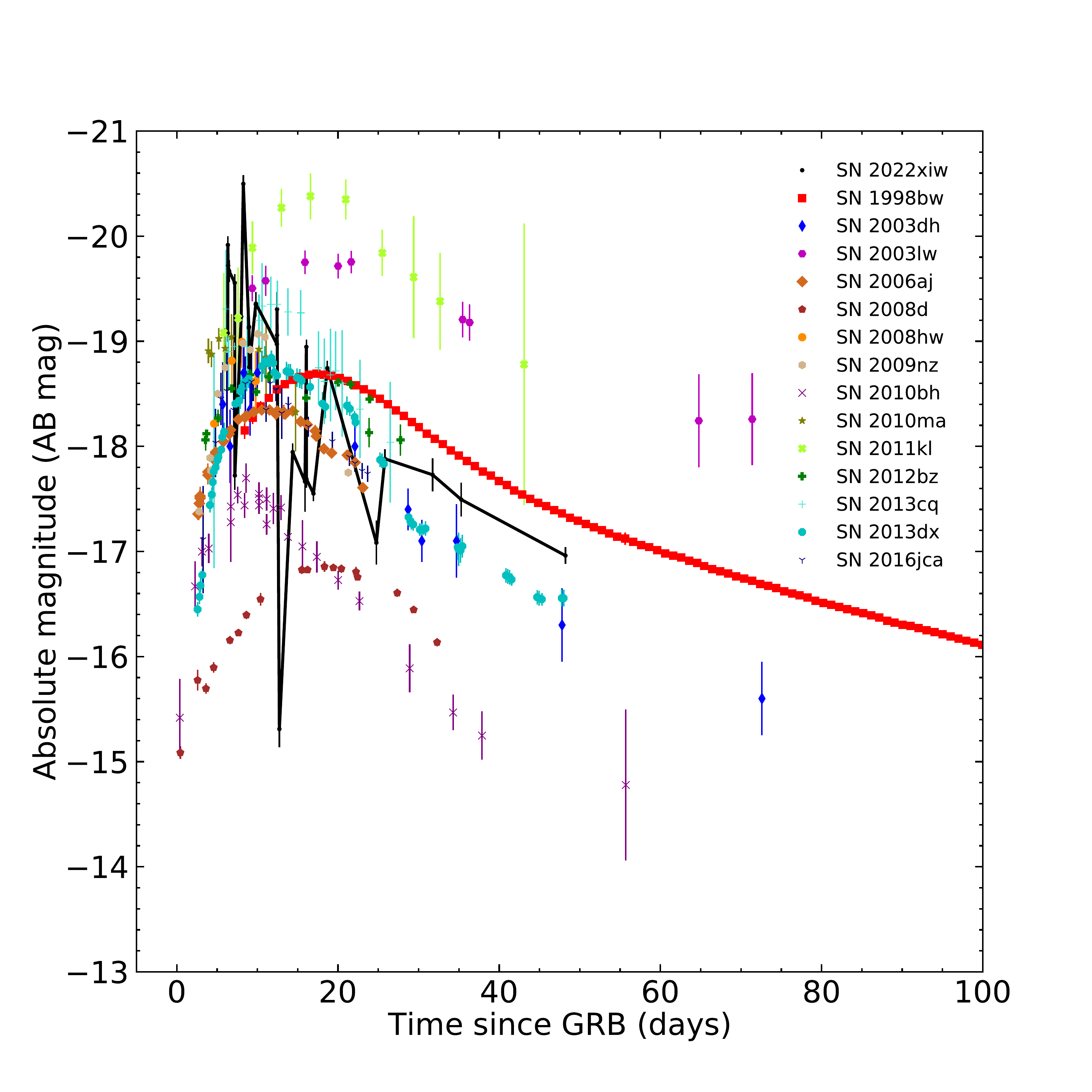}
\caption{The light curve of SN\,2022xiw (black), after subtracting the BPL and host-galaxy components. The data for other SNe are taken from \citep{2018ApJ...862..130L}.
\label{fig:AB_mag}}
\end{figure}

\begin{figure}[ht!]
\centering
\includegraphics[angle=0,scale=0.17]{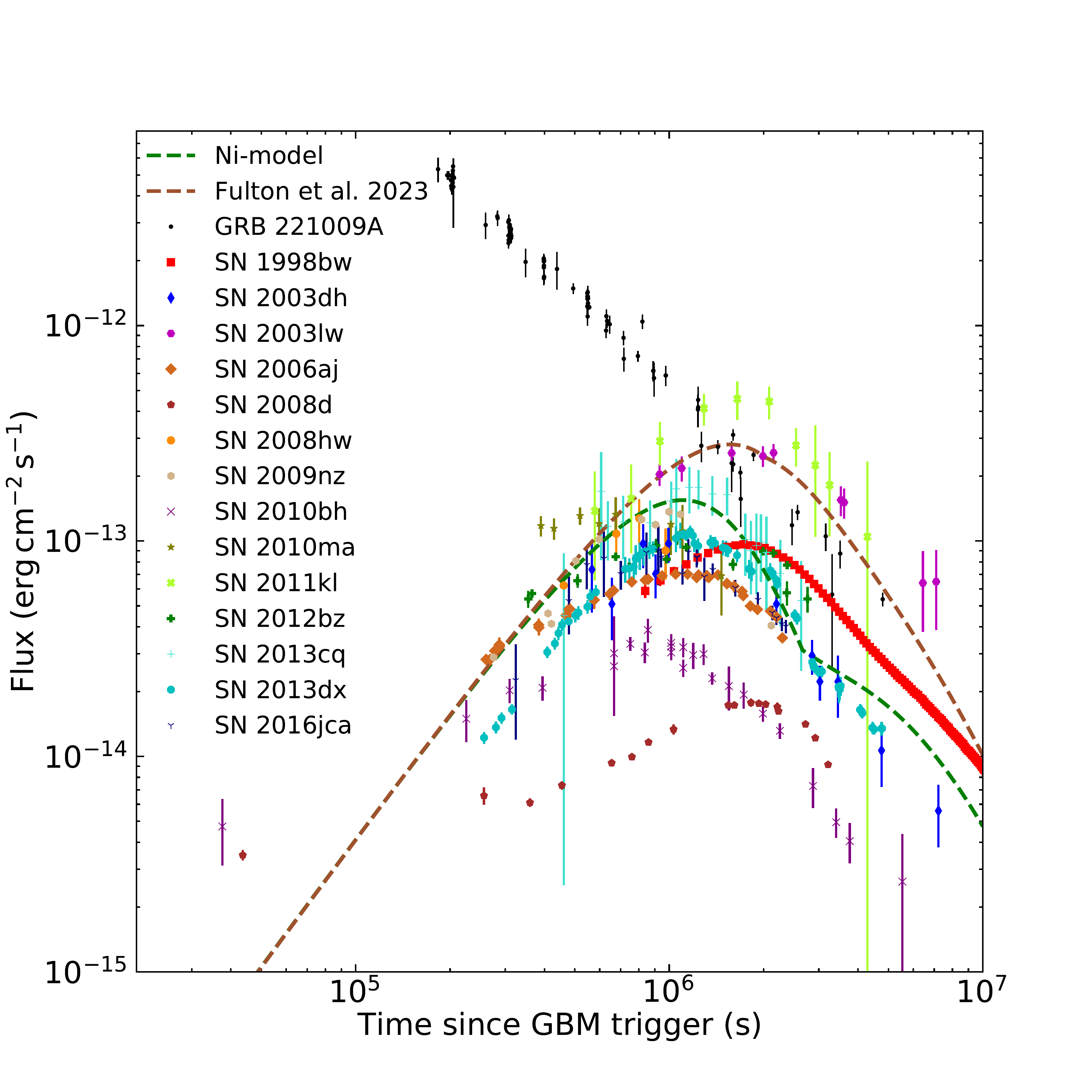}
\caption{Comparison of SN light curves at the same redshift ($z=0.151$), which shows that only two SNe (SN\,2003lw and SN\,2011kl) can be obviously found in the afterglow of GRB\,221009A.
\label{fig:SN-peak}}
\end{figure}

To study the pure SN light curve of SN\,2022xiw, we chose the $R$-band data from 7 days after the GBM trigger, then subtracted the BPL and host-galaxy components in our model, and finally discarded the negative values. We converted the data to absolute magnitude, which are shown in Figure \ref{fig:AB_mag}, though with large scatter.

To better compare among SNe, we set the other SN redshifts to be the same as that of GRB\,221009A, with the results shown in Figure \ref{fig:SN-peak}.
We find that only two SNe can be obviously detected in the afterglow of GRB\,221009A --- SN\,2003lw and SN\,2011kl. Most of the other SNe are obscured by the afterglow of GRB\,221009A. 
In our sample, SN\,2003lw has the largest peak time ($t_\mathrm{peak} = 21.5 \pm 3.5\,\mathrm{days}$) and SN\,2011kl is the brightest (with $M_\mathrm{peak} = -19.8\pm0.1$\,mag); indeed, SN\,2011kl resemble a superluminous SN \citep{2015Natur.523..189G,2019A&A...624A.143K}. 
Therefore, we suggest that an SN may be detected when its peak emerges relatively late, when the afterglow of the GRB has faded to a low level. Another possibility is that an SN is very luminous, in order to be found in the afterglow of a GRB.

\section{Conclusion and Discussion}\label{sec:Conclusion and discussions} 

We have presented multiwavelength observations of the extraordinary GRB\,221009A, spanning about 19\,days in time. 
A weak bump, possibly from a supernova, emerges from the declining afterglow at $\sim 11$\,days in the $R$-band light curve. Here we summarize our results.

\begin{enumerate}

\item We used a smooth broken power law plus $^{56}\mathrm{Ni}$ model to fit the $R$-band light curve. The best-fitting results reveal that the SN ejected a total mass of $M_\mathrm{ej} = 3.70\, M_\odot $, a $^{56}\mathrm{Ni}$ mass of $M_\mathrm{Ni} = 0.23 \, M_\odot$, and a total kinetic energy of $E_\mathrm{SN,K} = 2.35 \times 10^{52}\,\mathrm{erg}$. We estimate the energy partition $\eta = 0.11$.
In addition, we used a single-decline-rate power law, $f(t) \propto t^{-1.56}$, to describe multiband light curves, and found no significant SN signal in all bands.   

\item We compared GRB\,221009A with other GRB-SN events based on a GRB-associated SN sample, finding that the correlations among SN parameters of GRB\,221009A are consistent with other GRB-SN events. It is noteworthy that both the $\gamma$-ray energy and SN kinetic energy are large. We set these SNe in our sample at the distance of GRB\,221009A, and find that only SN\,2003lw and SN\,2011kl can be obviously detected in the afterglow of GRB\,221009A.

\end{enumerate}

Owing to the limited dataset and the bright afterglow, it is difficult to study the SN signals and give good constraints.
Focusing on our fitting results, on the one hand, 
we estimate the absolute AB mag of the SN peak to be $M_R = -19.21$\,mag, close to the value of $M_r=-19.4\pm0.3$\,mag that \cite{2023ApJ...946L..22F} measured, and also consistent with the limit $M_r>-19.5$\,mag that is given by \cite{2023ApJ...946L..25S}. 
On the other hand, \cite{2022GCN.32800....1D} announced spectroscopic detection of emerging SN features $\sim 8$\,days after the burst. \cite{2022GCN.32850....1M} also reported spectroscopic confirmation of an SN in LBT spectra at $t \approx 8.56$\,days.
Our peak time of $(1+z)t_\mathrm{peak}=12.73$ days is close to these values.
In general, at the peak brightness of the SN, its spectral features are most obvious.
\cite{2023ApJ...946L..25S} suggested that the absolute magnitude of the associated SN would have to be brighter than $M_r =  -20.66$\,mag after correcting for the extinction of $E(B-V) = 1.32$\,mag, for them to detect SN bumps during that time period. But from our light-curve analysis, we suggest a weak bump emerges from the declining afterglow during that time period, and it is most visible at $t \approx 11$\,days.

In addition, \cite{2023ApJ...946L..25S} discussed that it is possible an associated SN was below their detection limit if they consider the high extinction.
Although \cite{2023ApJ...946L..28L} reported that they do not see significant evidence for SN emission in their observations, they do not discard the possibility that an event somewhat less luminous than SN\,1998bw (and perhaps somewhat faster evolving or bluer) could simply have evaded detection in their observations. \cite{2023ApJ...946L..28L} also reported that the optical to mid-infrared (0.6--12\,$\mu$m) SED shows little evidence for variability from early to late times (0.5--55\,days), but late-time {\it JWST} observations show that the G140M+G235M spectrum significantly differs from a power-law continuum 13\,days after the burst \citep{2023GCN.33676....1B}. The late-time spectroscopy and photometry are well described by an SN and power-law afterglow. The close match with SNe~Ic-BL in particular demonstrates the presence of a typical GRB-SN in the spectrum. These observations provide the first clear detection of an SN associated with GRB\,221009A \citep{2024NatAs.tmp...65B}.
Moreover, \cite{2023ApJ...949L..39S} also announced that they found moderate evidence for the presence of an additional component arising from an associated SN and found that it must be substantially fainter than SN\,1998bw \citep[see more details in the review by][]{2023ApJ...949L..39S}.
Thus, our $R$-band light light is likely to contain contributions from an SN \citep[named SN\,2022xiw;][]{2022TNSCR3047....1P}.  

Both the light-curve analysis and comparative analysis imply that the SN is not faint. However, why can we not obviously detect the SN signal? 
\cite{2023ApJ...946L..25S} mentioned that the nondetection of an SN from GRB\,221009A may be because most of the energy is carried by the relativistic jet, not the bulk ejecta. But we estimated the energy partition $\eta$ to be only 0.032, consistent with that inferred for most other GRB-SNe. 
Therefore, we suggest that the energy partition is not the primary reason why we cannot detect significant SN signal.

When we set the redshift of SNe in our sample to be the same as that of GRB\,221009A, we find that most of them are obscured by the afterglow of GRB\,221009A. Only two SNe can be obviously detected: SN\,2003lw with $t_\mathrm{peak} = 21.5\pm3.5\,\mathrm{days}$, and SN\,2011kl with $M_\mathrm{peak} = -19.8\pm0.1$\,days. So, we suggest that an  SN can be detected in the GRB afterglow in GRB-SN events either if it appears very late (when the afterglow has faded) or is very luminous. In the case of GRB\,221009A, the extraordinarily bright afterglow is likely a reason why the SN was not detected by several observers.

In conclusion, we suggest that SN\,2022xiw is possibly associated with GRB\,221009A, but the SN emission is largely obscured by the afterglow of GRB\,221009A. Our analysis shows that the SN is bright, but the GRB afterglow is even brighter.

\section{Acknowledgments} \label{sec:ack}

This work is supported by the National Natural Science
Foundation of China (grant Nos. 12373042, U1938201, 12133003 and 12273005), China Manned Spaced Project (CMS-CSST-2021-B11), the Bagui Scholars Programme (W.X.-G.) and the Guangxi Science Foundation the National (grant No. 2023GXNSFDA026007).
A.V.F.'s group at U.C. Berkeley is grateful for financial assistance
from the Christopher R. Redlich Fund,
Gary and Cynthia Bengier, Clark and Sharon Winslow, Alan Eustace (W.Z. is a Bengier-Winslow-Eustace Specialist in Astronomy), and many other donors.  KAIT and its ongoing
operation were made possible by donations from Sun Microsystems, Inc.,
the Hewlett-Packard Company, AutoScope Corporation, Lick Observatory,
the U.S. NSF, the University of California, the Sylvia \& Jim Katzman
Foundation, and the TABASGO Foundation. We thank the staff at Lick
Observatory for their assistance. Research at Lick Observatory
is partially supported by a generous gift from Google.




\bibliography{sample631}{}
\bibliographystyle{aasjournal}
\end{document}